\documentclass[twocolumn,prl]{revtex4}
\usepackage{epsf}
\usepackage{amsmath,amsthm,amssymb}
\usepackage{color}
\usepackage{dcolumn}
\usepackage{bm}
\usepackage{graphicx,epsfig}
\usepackage[dvipsnames]{xcolor}

\graphicspath{{fig/}}
\everymath{\displaystyle}

\begin{document} 

\title{Relativistic quantum-mechanical description of twisted
paraxial electron and photon beams}

\author{Alexander J. Silenko$^{1,2,3}$}
\email{alsilenko@mail.ru}
\author{Pengming Zhang$^{4}$}
\email{zhangpm5@mail.sysu.edu.cn}
\author{Liping Zou$^{1}$}
\email{zoulp@impcas.ac.cn}

\affiliation{$^1$Institute of Modern Physics, Chinese Academy of
Sciences, Lanzhou 730000, China}
\affiliation{$^2$Bogoliubov Laboratory of Theoretical Physics, Joint
Institute for Nuclear Research, Dubna 141980, Russia}
\affiliation{$^3$Research Institute for Nuclear Problems, Belarusian
State University, Minsk 220030, Belarus}
\affiliation{$^4$School of Physics and Astronomy, Sun Yat-sen University, Zhuhai 519082, China}

\date{\today}

\begin{abstract}
The analysis of twisted (vortex) paraxial photons
and electrons is fulfilled in the framework of relativistic quantum mechanics.
The use of the Foldy-Wouthuysen
representation radically simplifies the description of relativistic electrons and
clarifies fundamental properties of twisted particles. It is demonstrated that
the twisted and other structured photons are luminal.
Their subluminality apparently takes place because the photon energy is also contributed
by a hidden motion. This motion is vanished by averaging and disappears in the semiclassical description based on expectation values of the momentum and velocity operators. It is proven that semiclassical quanta of structured light are subluminal and massive. The quantum-mechanical and semiclassical descriptions of twisted and other structured electrons lead to similar results.
The new effect of a quantization of the velocity and the effective mass of the structured photon and electron is predicted. This effect is observable
for the photon. The twisted and untwisted semiclassical photon and electron modeled by the centroids are
considered in the accelerated and rotating noninertial frame. The coincidence of their inertial masses
with kinematic ones is shown. The orbital magnetic moment of the
Laguerre-Gauss electron does not depend on the radial quantum number.
\end{abstract}

\maketitle

The prediction and discovery of twisted (vortex) states of
photons \cite{Allen,He} and electrons \cite{Bliokh2007,UTV} has
opened new horizons in contemporary physics. In these states,
photons and electrons have orbital angular momenta (OAMs). At present, twisted photon and
electron beams are objects of intensive studies and have many practical applications (see Refs.
\cite{TwPhotons1,TwPhotons2,TwPhotons3,TwPhotons4,TwPhotons5,TwPhotRev1,TwPhotRev2,TwPhotRev3,
TwPhotRev4,TwPhotRevadd,TwPhotRevPadgett,BliokhPhotPhysRep,HarrisGrillo,BliokhSOI,Lloyd,LarocqueTwEl}
and references therein). The most important kind of such beams
is a paraxial (Laguerre-Gauss) wave beam
\cite{Allen,BliokhSOI,Siegman}
satisfying the paraxial approximation ($p_\bot<<p$). This beam
is unlocalized in the longitudinal direction $z$ and transversely 2D-localized. It is
described by the paraxial equation:
\begin{equation}
\begin{array}{c}
\left(\nabla^2_\bot+2ik\frac{\partial}{\partial
z}\right)\!\Psi=0,\quad
\nabla^2_\bot=
\frac{\partial^2}{\partial r^2}+\frac1r\frac{\partial}{\partial
r}+\frac{1}{r^2}\frac{\partial^2}{\partial\phi^2}.
\end{array}
\label{eqp}
\end{equation}
The known solutions in cylindrical coordinates
are the Laguerre-Gauss beams \cite{Allen,BBP,PlickKrenn}:
\begin{equation}
\begin{array}{c}
\Psi={\cal A}\exp{(i\Phi)},\\
{\cal A}=\frac{C_{nl}}{w(z)}\left(\frac{\sqrt2r}{w(z)}\right)^{|l|}
L_n^{|l|}\left(\frac{2r^2}{w^2(z)}\right)\exp{\left(-\frac{r^2}{w^2(z)}\right)},\\
\Phi=l\phi+\frac{kr^2}{2R(z)}-(2n+|l|+1)\varphi(z),\\
C_{nl}=\sqrt{\frac{2n!}{\pi(n+|l|)!}},\quad w(z)=w_0\sqrt{1+\frac{4z^2}{k^2w_0^4}},\\
R(z)=z+\frac{k^2w_0^4}{4z},\quad
\varphi(z)=\arctan{\left(\frac{2z}{kw_0^2}\right)},
\end{array}
\label{eq33new}
\end{equation}
\begin{equation}
\int{\Psi^\dag\Psi rdrd\phi}=1,
\label{eqint}
\end{equation}
where the real functions ${\cal A}$ and $\Phi$ define the amplitude and phase of the beam,
$k$ is the beam wavenumber, $w_0$ is the minimum beam
width, $L_n^{|l|}$ is the generalized Laguerre polynomial, and
$n = 0, 1, 2,
\dots$ is the radial quantum number. A frequently encountered inexactness
\cite{BliokhSOI,Lloyd,LarocqueTwEl} is the superfluous factor $\exp{(ikz)}$.
Other quantum-mechanical solutions are 3D-localized particle wavepackets
\cite{BliokhSOI,Voloch-Bloch,Zhongetal,KarlovetsGaussian,FDengDDeng,KarlovetsScattering,KarlovetsParaxial}.
Quantum numbers of twisted photons have been determined in Ref. \cite{BialynickiBirulaPhysScr}.

We assume that $\hbar=1,~c=1$ but include $\hbar$ and
$c$ into some formulas when this inclusion clarifies the
problem.

One of the most mysterious physical phenomena is the recently discovered subluminality of free twisted photons for Bessel \cite{Giovannini} (see also Ref. \cite{AlfanoNolanBessel}) and Laguerre-Gauss \cite{Bouchard,Lyons} beams. Special relativity asserts that massless particles in vacuum should be luminal. Therefore, a definite solution of this puzzle should be based on a description of single photons. In the present work, we make this description in the framework of relativistic quantum mechanics (QM). We investigate new properties of twisted particles and untwisted but structured ones changing a usual perception of such
particles.

The possibility to use a quantum-mechanical approach for a description of light quanta is nontrivial and should be substantiated. For the photon in optics, $\Psi$ is not a wave function in the same sense as for the electron and is simply a function that determines the relative amplitude of the electric field \cite{Allen,Siegman}. 
The full description of an electromagnetic field including its interaction with matter is based on the quantum field theory (see Refs. \cite{Loudon,Srednicki}). However, the propagation of light in a free space can be adequately described with the Riemann-Silberstein vector $$\bm F=\frac{1}{\sqrt2}\left(\bm E+ i\bm B\right).$$ It 
allows one to reduce the Maxwell equations and to present them in the form
\cite{BialynickiBirulaPhysScr,Dirac-like}
\begin{equation}
i\hbar\partial_t\bm F=c(\bm\tau\cdot\bm p)\bm F, \label{Weyllike}
\end{equation} where $\bm\tau$ is a vector in which the components are the conventional spin-1 matrices acting on three components of $\bm F$. This equation is similar to the Weyl equation for a massless spin-1/2 neutrino \cite{Dirac-like}.
When the six-component wave function is defined by \cite{Barnett-FWQM}
\begin{equation}
\psi=\frac{1}{\sqrt2}\left(\begin{array}{c} \phi \\ \chi \end{array}\right),\quad \phi=\left(\begin{array}{c} E_x \\ E_y \\ E_z \end{array}\right),\quad \chi=\left(\begin{array}{c} iB_x \\ iB_y \\ iB_z \end{array}\right), \label{wavefnct}
\end{equation}  the Dirac-like equation for the free electromagnetic field can be obtained:
\begin{equation}
i\hbar\frac{\partial\psi}{\partial t}=\bm\alpha\cdot\bm p\,\psi, \qquad\bm\alpha=\left(\begin{array}{cc} 0 & \bm\tau \\ \bm\tau & 0 \end{array}\right).\label{Diraclike}\end{equation} In this connection, we can mention the existence of bosonic symmetries of the standard Dirac equation \cite{Simulik,Simulik1,Simulik2,Simulik3,Simulik4,Simulik5}.
For twisted paraxial photons and electrons, we use the Foldy-Wouthuysen (FW) representation \cite{FW} in relativistic QM obtained by appropriate unitary transformations of initial Hamiltonians and wave functions. Wonderful advantages of this representation are restoring the Schr\"{o}dinger form of relativistic QM and unifying
relativistic QM for particles with different spins \cite{FW,FeshbachVillarsetal,unification}. In the FW representation,
the Hamiltonian and all fundamental operators are block-diagonal
(diagonal in two spinors or spinor-like parts of wave functions). The passage to the classical limit
usually reduces to a replacement of the operators in quantum-mechanical
Hamiltonians and equations of motion with the
corresponding classical quantities \cite{JINRLett12}. The FW wave function being a generalization of the Schr\"{o}dinger wave function on the relativistic case
permits the probabilistic interpretation \cite{Reply2019}. Thanks to these properties, the FW
representation provides the best possibility to obtain a
meaningful classical limit of relativistic QM not only for the
stationary case \cite{FW,JINRLett12,CMcK,JMPcond} but also for the
nonstationary one \cite{PRAnonstat}.

We use the results obtained in Ref. \cite{Barnett-FWQM}. The FW transformation of the Dirac-like Hamiltonian $\bm\alpha\cdot\bm p$ is straightforward and the FW Hamiltonian for a free photon is defined by
\cite{Barnett-FWQM}
\begin{equation}
{\cal H}_{FW}\Psi_{FW}=\beta|\bm p|\Psi_{FW},\quad \beta={\rm
diag}(1,1,1,-1,-1,-1).
\label{FWphoton}
\end{equation}
While the wave functions $\Psi$ and $\Psi_{FW}$ have different definitions, a connection between $\bm E$ and $\bm B$ provides for their similarity. $\Psi_{FW}$ is anyway proportional to a field amplitude. For a plane electromagnetic wave, $\bm B=\bm p\times\bm E/p$. Importantly, the quantum-mechanical approach allows one to introduce operators and to calculate their expectation values.

The corresponding FW Hamiltonian for a free electron \cite{FW} is similar to that for the free photon:
\begin{equation}
{\cal H}_{FW}=\beta\sqrt{m^2+\bm p^2},\quad \beta={\rm
diag}(1,1,-1,-1).
\label{FWelectron}
\end{equation}
FW Hamiltonians describing free massive spin-1 particles \cite{Case,spinunit} and
massive and massless scalar ones \cite{FeshbachVillarsetal} are also similar. The number of components of corresponding wave functions depends on the spin and is equal to $2(2s+1)$.

Despite the similarity of the FW Hamiltonians, the wave functions for the photon and electron substantially differ from each other. The photon wave function $\Psi_{FW}$ characterizes
the relative amplitude of the electromagnetic field \cite{Barnett-FWQM} and cannot be regarded as the probability
amplitude of the spatial localization of the photon (\cite{LL4}, p. 12). As a contrary, the corresponding wave function for the electron enables the probabilistic interpretation \cite{Reply2019}. However, the
photon wave function defines 
eigenvalues or expectation values of all operators. Its squared magnitude, $|\Psi_{FW}|^2$, is proportional to the light energy density. The physical reality
of wave functions of twisted photons has been confirmed in Ref. \cite{PhysicalReality}.

Lower spinors or spinor-like parts of
FW wave functions vanish \cite{vanishing}. Hereinafter, they will be eliminated and $\beta$ matrices will be removed.
The simple form of Eqs. (\ref{FWphoton}) and (\ref{FWelectron}) clearly shows preferences of the
approach based on the FW transformation.

The standard quantum-mechanical approach based on the Proca equations brings a result which is in accordance with Eq. (\ref{FWphoton}). For massive and massless free spin-1 particles, these equations lead to the following second-order equation (\cite{LL4}, Eq. (14.4)):
\begin{equation}
(p_0^2-\bm p^2-m^2)\psi_\mu=0,\quad p_0\equiv i\frac{\partial}{\partial t},
\label{eqscord}
\end{equation}
where $\psi_\mu~(\mu=0,1,2,3)$ has three \emph{independent}
components. For the photon, $m=0$ and Eqs. (\ref{FWphoton}) and (\ref{eqscord}) agree.

Optical and quantum-mechanical approaches 
significantly differ.  
Optics studies the light field and determines its \emph{local} velocities.
Certainly, phase and group velocities are different.  A local phase velocity (LPV) is
defined by the phase front $\Phi(\bm r)$, $v_p=\omega/|\nabla\Phi(\bm r)|$,
where $\omega=ck$ is the angular frequency \cite{Born-Wolf,Bouchard}.
Another frequently used formula for the LPV has been obtained in Ref. \cite{Chen}
(see also Ref. \cite{HuangWuHu}):
\[
v_p=c\left[1+\frac{\nabla^2{\cal A}(\bm r)}{k^2{\cal A}(\bm r)}\right]^{-1/2}.
\]
The local group velocity is given by $v_g=|\partial_\omega\nabla\Phi(\bm r)|^{-1}$
\cite{Born-Wolf,Bouchard} (see also Ref. \cite{Saari} for details).
The analysis shows
\cite{Bouchard,HuangWuHu,Saari,Lunardi,WangScheidHo,Qian,ZhangWangXie,Vasnetsov,WangWang,Major,Garay-Avendano,Kondakci}
that the both velocities can be subluminal and superluminal depending on a region.
Certainly, both the local phase and group velocities characterize important properties of
twisted light beams. For example, the LPV defines an electron acceleration in a
laser beam \cite{WangScheidHo,ZhangWangXie,WangHoYuan}. The distribution of the LPV
has been measured in Ref. \cite{WangQian}.

However, any free photon at any time extends over the whole 3D-space and thus the optical approach based on the \emph{local} phase and group velocities
may fail to determine its fundamental properties (quantum numbers \cite{BialynickiBirulaPhysScr} and eigenvalues and expectation values of operators). As a contrary, the quantum-mechanical approach providing for a \emph{single-particle} description perfectly solves this problem.

For stationary states (${\cal H}_{FW}\Psi_{FW}=E\Psi_{FW}$), squaring
Eq. (\ref{FWelectron}) for the upper spinor and applying the paraxial approximation for
$p_z>0$ results in (cf. Ref. \cite{Barnett-FWQM})
\begin{equation}
p=\sqrt{p_\bot^2+p_z^2}\approx p_z+\frac{p_\bot^2}{2p},\qquad p=\hbar k=\sqrt{E^2-m^2}.
\label{eqnnumb}
\end{equation}
The operator form of Eq. (\ref{eqnnumb}) reads
\begin{equation}
\begin{array}{c}
\left(\nabla^2_\bot+2ik\frac{\partial}{\partial
z}\right)\Psi_{FW}=-2k^2\Psi_{FW}.
\end{array}
\label{eqpqm}
\end{equation}
The substitution $\Psi_{FW}=\exp{(ikz)}\Psi$ brings the paraxial equation (\ref{eqp}).
Within the paraxial approximation, it \emph{exactly} describes photons and electrons of
arbitrary energies. Therefore, the FW transformation radically simplifies a
description of relativistic electrons (cf. Ref. \cite{BialynickiBirulaPRL}). We
underline the difference between $\Psi_{FW}$ and $\Psi$.

The subluminality of twisted (and untwisted) light finds a straightforward
explanation and description in relativistic QM which is a part of quantum optics. All beam parameters are defined
by expectation values or eigenvalues of related operators. QM shows that the
twisted photon is luminal
and its subluminality is \emph{apparent}. The group velocity operator,
$v\equiv\sqrt{v_r^2+v_\phi^2+v_z^2}$, depends on a hidden motion 
in the horizontal plane \cite{footnote}.  As follows from Eq. (\ref{FWelectron}),
\begin{equation}
\begin{array}{c}
\bm v=\frac{\partial{\cal H}_{FW}}{\partial\bm p}=\frac{c\bm p}{p},\quad v=c.
\end{array}
\label{velocity}
\end{equation} We use the term ``hidden motion'' for a motion which does not contribute to expectation values of operators defining some components of the velocity and momentum but affects both expectation values of squares of these operators and eigenvalues of the energy operator. In the considered case, expectation values of two Cartesian velocity components are zero ($<v_i>=0,~i=x,y$). However, $<v_x^2+v_y^2>=<v_r^2+v_\phi^2>\neq0$. Importantly, just expectation values of main operators define \emph{measurable} beam parameters.
For the electron, the velocity operator reads
\begin{equation}
\begin{array}{c}
\bm v=\frac{c\bm p}{\sqrt{m^2c^2+\bm p^2}}.
\end{array}
\label{veloe}
\end{equation}

Certainly, only the $z$ component of the group velocity $\bm v$ can be \emph{directly}
measured. For the photon, it is less than $c$. This fact creates the impression that the
twisted photon is subluminal.

QM is a foundation of contemporary physics and measurable quantities are expectation values of the velocity and momentum operators. Therefore, the \emph{classical} model of light quanta (Einstein quanta) which velocity, energy, and momentum are defined by expectation values or eigenvalues of the corresponding operators remains very important. For twisted and any other structured light, the result is nontrivial.

The calculation of expectation values of $v_z$ is straightforward. It follows
from Eqs. (\ref{eqp}), (\ref{eqnnumb}), and (\ref{velocity}) that
\begin{equation}
\begin{array}{c}
\frac{v_z}{c}=\sqrt{1-\frac{p_\bot^2}{p^2}}
=\sqrt{1-\frac{2i}{k}\frac{\partial}{\partial z}}\approx1-\frac{i}{k}\frac{\partial}{\partial z}.
\end{array}
\label{veloz}
\end{equation}
As ${\cal A}^\dag={\cal A},~{\cal A}^2=|\Psi|^2$,
\begin{equation}
\begin{array}{c}
\int{\!\Psi^\dag\frac{\partial\Psi}{\partial z} rdrd\phi}=\int{\!{\cal A}\frac{\partial{\cal A}}{\partial z} rdrd\phi}+i\int{\!|\Psi|^2\frac{\partial\Phi}{\partial z} rdrd\phi}.
\end{array}
\label{eqf}
\end{equation}
The first integral in the right-hand side vanishes:
\[
\int{{\cal A}\frac{\partial{\cal A}}{\partial z} rdrd\phi}=\frac12\frac{d}{dz}\int{|\Psi|^2rdrd\phi}=0.
\]
The second integral can be calculated exactly. Since
\begin{eqnarray}
 \frac{\partial\Phi}{\partial z}
&=&\frac{2}{kw^2(z)}\left\{\frac{r^2}{w_0^2}\left[1-\frac{8z^2}{k^2w_0^2w^2(z)}\right]-\zeta\right\},\nonumber \\
 \zeta&=&2n+|l|+1,
\label{eqphi}
\end{eqnarray}
averaging (see Ref. \cite{integral}) results in
\begin{equation}
\begin{array}{c}
\left\langle r^2\right\rangle=\frac{\zeta w^2(z)}{2},\quad
\left\langle\frac{\partial\Phi}{\partial z}\right\rangle
=-\frac{\zeta}{kw_0^2},\quad \left\langle p^2_\bot\right\rangle
=\frac{2\zeta}{w_0^2},
\end{array}
\label{eqpha}
\end{equation}
\begin{equation}
\begin{array}{c}
<v_z>=c\left(1+\frac{1}{k}\left\langle\frac{\partial\Phi}{\partial z}\right\rangle\right)
=c\left(1-\frac{2n+|l|+1}{k^2w_0^2}\right).
\end{array}
\label{eqphf}
\end{equation}
A comparison of Eqs. (\ref{eqphi}) and (\ref{eqphf}) shows that the contributions
from regions with small and large values of $r$ to $v_z$ are subluminal and superluminal, respectively.

Equation (\ref{eqphf}) has been previously derived in Ref. \cite{subluminalBarezaHermosa}. However, the right interpretation of this equation can be based only on relativistic QM. Our approach connects the result (\ref{eqphf}) with initial quantum-mechanical equations (\ref{FWphoton}) and (\ref{eqscord}) and, therefore, attributes it to a \emph{single} photon. All twisted and untwisted Laguerre-Gauss modes, including the fundamental mode $n=l=0$, are subliminal.
Our results do not support the formula obtained in Ref. \cite{Tamburini} by averaging the
local field velocity which does not characterize the single photon. For the electron, Eq. (\ref{eqpha}) remains unchanged and the longitudinal
velocity is given by
\begin{equation}
\begin{array}{c}
<v_z>=\frac{ck}{\sqrt{k^2+K^2}}\left(1-\frac{2n+|l|+1}{k^2w_0^2}\right),\quad K=\frac{mc}{\hbar}.
\end{array}
\label{eqphfel}
\end{equation}

We predict the new property of twisted particles consisting in a quantization of the group
velocity and following from Eqs. (\ref{eqphf}) and (\ref{eqphfel}). We suppose that this
quantization can be observed because the modes $n$ and $l$ are measurable \cite{WangLiChen}. Experimental data \cite{Bouchard} obtained for mixtures of modes with different $n$ agree with our prediction but cannot prove it.

Some properties of twisted particles characterize a local field while other properties
are attributed to the photon or electron extending over the whole spacetime.
In particular, $<r^2>$ depends on $z$ and depicts local field properties.
As a contrary, $<p_\bot^2>$ and $<v_z>$ are independent of $z$ and define general
quantum-mechanical parameters of the twisted photon and electron.

Since wave properties of twisted particles are defined by $<p_z>$ and $<v_z>$ and a
detailed analysis of the hidden transversal motion can often be avoided, it is
convenient to consider such particles as extended 
objects (the so-called centroids \cite{BliokhPhotPhysRep,BliokhSOI}) moving
in the $z$ direction. This model remains applicable for twisted particles in external fields \cite{BliokhPhotPhysRep,BliokhSOI,Manipulating,ResonanceTwistedElectrons,PhysRevLettEQM2019,snakelike}. The transition to the semiclassical approximation allows
us to determine mechanical properties of the centroids. In this case, angular brackets
can be omitted and we can consider a twisted photon like a centroid with the constant
lab frame energy $E=\sqrt{p_z^2+p_\bot^2}$. The velocity of the centroid is defined
by $v_x=v_y=0,~v_z=p_z/E$ and the origin of an internal motion defined by $p_\bot^2$
can be disregarded. Certainly, such a quasiparticle satisfies the requirements of special
relativity only if it possesses the mass $M=\sqrt{E^2-p_z^2}$.

The validity of introduction of the light mass was previously studied only for
\emph{groups of photons}. It is known \cite{Okun} that two photons with equal
frequencies and with the angle $2\theta$ between the directions of their wave vectors
acquire the Lorentz-invariant mass $m=(2\hbar\omega/c^2)\sin{\theta}$.
In Refs. \cite{FedorovLaserPhys,FedorovIOP}, this property has been applied to
groups of \emph{nonidentical and noncollinear} photons containing Gaussian pulses.
It has been underlined \cite{FedorovLaserPhys,FedorovIOP} that such approach is
inapplicable for single photons or groups of identical photons being objects of
our study. We can add that the Gaussian pulses  describe neither twisted states
nor untwisted states with a nonzero radial quantum number. In particular, the
average velocity obtained in Refs. \cite{FedorovLaserPhys,FedorovIOP,FedorovEPL}
reads $<v_z>=c[1-(2k^2w_0^2)^{-1}]$ (cf. Eq. (\ref{eqphf})).

To verify a possibility to model the Laguerre-Gauss photon by a massive centroid,
we need to pass to an arbitrary inertial frame.
Let us make the Lorentz boost to the centroid rest frame ($v_z^{(0)}=0$). In this
frame, $E^{(0)}\!=\!\sqrt{p_\bot^2}, ~ p_x^{(0)}\!=\!p_x\!=\!0,~ p_y^{(0)}\!=\!p_y\!=\!0,~ p_z^{(0)}\!=\!0$.
We can now consider the second boost to the frame denoted by primes and moving
with the arbitrary velocity $-\bm V$ relative to the centroid rest frame. If we
change the coordinates and direct the $X$ axis along the vector $\bm V$, the
Lorentz transformation results in
\begin{equation}
\begin{array}{c}
p_X'=\frac{VE'}{c^2},\quad p_Y'=p_Z'=0,\quad
E'=\frac{E^{(0)}}{\sqrt{1-\frac{V^2}{c^2}}}.
\end{array}\label{Ltran}
\end{equation}
It is easy to check that arbitrary Lorentz transformations for the centroid are
equivalent to those for a massive particle with the mass $M=E^{(0)}/c^2$.
When $\hbar,c$ are included, the effective mass of the twisted photon (centroid mass) reads
\begin{equation}
\begin{array}{c}
M=\frac{\sqrt{2(2n+|l|+1)}\hbar}{cw_0}.
\end{array}
\label{masseff}
\end{equation}
Its relation to the centroid velocity is defined by
\begin{equation}
\begin{array}{c}
M=\frac{\hbar k}{c}\sqrt{2\left(1-\frac{<v_z>}{c}\right)}
=\frac{\hbar k}{c}\sqrt{1-\frac{<v_z>^2}{c^2}}.
\end{array}
\label{massphv}
\end{equation}
This result shows a nontrivial possibility of
conversion of the Lorentz-noninvariant hidden momentum into the Lorentz-invariant mass.
The mass-energy ratio is given by
\begin{equation}
\begin{array}{c}
\frac{Mc^2}{E}=\frac{\sqrt{2(2n+|l|+1)}\lambda}{2\pi w_0},\quad
\lambda=\frac{2\pi}{k}.
\end{array}
\label{massenergy}
\end{equation}

The second boost, unlike the first one, changes the OAM \cite{Manipulating}.

We can easily extend our analysis to the other forms of structured light. Equations (\ref{FWphoton}) and (\ref{eqscord}) remain valid in the general case. Our derivation 
covers Gaussian beams because the presence or absence of the OAM is not important in this case. The other forms of structured light are also characterized by the hidden motion. For the 3D-localized particle wavepackets (light bullets) \cite{FDengDDeng,Abdollahpour,Zhong-Airy}, wave functions are 3D-normalized ($\int{\Psi^\dag\Psi d^3x}=1$) and this motion takes place in three directions. The Lorentz boost from the wavepacket rest frame to the lab frame also satisfies Eq. (\ref{Ltran}) for any chosen direction $X$. In this case, $E^{(0)}=\sqrt{\bigl(p_x^{(0)}\bigr)^2+\bigl(p_y^{(0)}\bigr)^2+\bigl(p_z^{(0)}\bigr)^2}$. Thus, arbitrary Lorentz transformations for the light wavepacket are
equivalent to those for a massive particle with the mass $M=E^{(0)}/c^2$. The relation (\ref{massphv}) also remains unchanged. Equations (\ref{FWphoton}) and (\ref{eqscord}) demonstrate that the velocity operator is equal to $c$ for any form of light. For wavepackets, one can also determine the parameters of semiclassical light quanta (Einstein quanta) by averaging the momentum and velocity operators. Evidently, such semiclassical quanta are subluminal and massive.

To complete the analysis, we need only to consider Laguerre-Gauss and other structured particles in
noninertial frames. This consideration allows us to determine an \emph{inertial mass}
which is important in processes of beam acceleration and rotation. The light
beam acceleration and rotation are largely investigated (see Refs. \cite{acceleration,rotation}
and references therein). The problem is rather nontrivial. In particular,
the kinematic (``Lorentz-invariant'' \cite{FedorovLaserPhys,FedorovIOP,FedorovEPL})
mass of the group of noncollinear photons may not manifest itself in inertial
and gravitational interactions \cite{Okun,FedorovLaserPhys,FedorovIOP,FedorovEPL}.
Practical importance of the related problem of Laguerre-Gauss photons in
gravitational fields is not so great.

For spinning and spinless single particles in noninertial frames, relativistic
FW Hamiltonians and equations of motion as well as their classical counterparts
have been derived in Refs. \cite{PhysRevD,ostrong,gravityDirac,gravityspinzero,gravityProca}.
We may disregard spin effects because corresponding terms in the Hamiltonians
are relatively small. In the semiclassical approximation, the Hamiltonian of a
particle in a noninertial frame accelerated with the acceleration $\bm a$
and rotating with the angular velocity $\bm\omega$ has the form \cite{ostrong}
\begin{equation}
\begin{array}{c}
{\cal H}=\left(1+\bm a\cdot\bm
r\right)
\sqrt{m^2+\bm p^2}-\bm\omega\cdot\bm l.
\end{array}\label{Hamltni}
\end{equation}
Here $\bm a$ and $\bm\omega$ are
independent of the spatial coordinates but may arbitrarily depend on time \cite{ostrong}
and ${\bm l}$ is the total angular momentum. The particle motion is affected by
the accelerator, Coriolis and centrifugal forces. If sizes
of the light beam are negligible as compared with those of the beam trajectory
in the inertial field, $\bm l=\bm r\times\bm p+\bm
L$, where $\bm L$ is the intrinsic OAM. For the Laguerre-Gauss light beam ($m=0$) formed by
identical photons, the semiclassical approximation consists in
$\bm p^2\rightarrow p_z^2+p_\bot^2,~\bm l\rightarrow (\bm r\times \bm e_z)p_z+\bm L$.
The $z$ axis is longitudinal. Evidently, the paraxial photon should be modeled by
the \emph{massive} centroid with the \emph{inertial} mass $M$ defined by Eq. (\ref{masseff}).
Twisted and untwisted photons with the same energy have different momenta, velocities,
and Lorentz factors and can be distinguished. These conclusions remain valid for other structured photons (in particular, for light wavepackets).

The nonzero mass as well as the subluminal velocity are extraordinary properties
of the Laguerre-Gauss photon. The longitudinal beam shape depends on $z/z_R$,
where $z_R = kw^2_0/2$ is the Rayleigh diffraction length. The
last quantity, in particular, does not satisfies the Lorentz transformations
for a segment length. Therefore, the independence of centroid parameters
from $z$ is necessary to use the model of the centroid.

Despite the paraxial approximation $\left(<p_\bot^2>\ll p^2\right)$, the
Laguerre-Gauss photon mass is not very small. Under the experimental conditions
used in Ref. \cite{Bouchard}, $Mc^2/E\approx0.02$ when $\zeta$=100 and $E$=1.56 eV.

The presented consideration remains valid for the twisted electron. When the
hidden transversal motion is taken into account, the electron velocity $v$
satisfies Eq. (\ref{veloe}).
However, the twisted electron can also be regarded as the centroid with the
velocity $v_z$ given by Eq. (\ref{eqphfel}) and with the mass equal to
\begin{equation}
\begin{array}{c}
M=\sqrt{m^2+<p_\bot^2>}=\sqrt{m^2+\frac{2(2n+|l|+1)}{w_0^2}}.
\end{array}
\label{masselv}
\end{equation}
Amazingly, the relation between the mass and velocity of the Laguerre-Gauss
electron almost coincides with Eq. (\ref{massphv}):
\begin{equation}
\begin{array}{c}
M=\frac{E}{c^2}\sqrt{1-\frac{<v_z>^2}{c^2}}.
\end{array}
\label{masselt}
\end{equation}
The centroid momentum is equal to
\[
<p_z>=\sqrt{E^2-m^2}\left[1-\frac{2n+|l|+1}{(E^2-m^2)w_0^2}\right].
\]

For the paraxial electron in noninertial frames, the only difference
from the paraxial photon is the nonzero mass $m$.
The \emph{inertial} mass of the corresponding centroid is defined by
$M=\sqrt{m^2+p_\bot^2}$ and coincides with the kinematic mass (\ref{masselv}).

A similar effect of an increase of the kinematic (Lorentz-invariant)
mass of \emph{3D-localized wavepackets} of free twisted electrons as
compared with $m$ has been found in Ref. \cite{KarlovetsParaxial}.

Importantly, the effective masses of the twisted paraxial photon and
electron (i.e., the corresponding centroid masses) are quantized. The twisted wavepackets also possess this property. The
quantization of the mass and the group velocity can be discovered
simultaneously in view of Eqs. (\ref{massphv}) and (\ref{masselt}).

We underline that all Laguerre-Gauss beams, even the mode $n=l=0$, and all twisted and untwisted wavepackets move slower than the plane wave and have the mass $M>m$ ($v_g<c$ and $M>0$ for light).

Since the Laguerre-Gauss electron is charged, it possesses a magnetic
moment. Due to the connection between the FW operators of the OAM and
the orbital magnetic moment, $\bm\mu_L=e\bm L/(2E)$
\cite{Manipulating,Barut,BliokhPRL2011,Kruining,ResonanceTwistedElectrons},
the latter is not influenced by the radial quantum number.
The total magnetic moment contains also a spin part (see Refs. \cite{BMT,Fradkin-Good,PhysScr}).

In this Letter,
we have performed a general description of twisted paraxial photons
and electrons in the framework of relativistic QM. The use of the FW
representation has allowed us to find and investigate their new properties
changing the usual perception of such
particles. In this representation, twisted paraxial photons and electrons of
arbitrary energies are characterized by the well-known wave function (\ref{eq33new})
and, therefore, a description of relativistic electrons is radically simplified.
Moreover, the quantum-mechanical approach clarifies fundamental properties of
single photons and electrons. We have checked that twisted and other structured photons are luminal.
Their subluminality is \emph{apparent} and appears because the photon energy is
contributed by the hidden 
motion. For Laguerre-Gauss light beams, this motion is transversal and the average transversal momentum
vanishes. For light wavepackets, the hidden motion occurs in three directions. We have presented the quantum-mechanical and semiclassical descriptions of the structured photon. In QM, such photon is 
a massless particle moving with the velocity
$c$. The semiclassical description applies expectation values of the momentum and velocity operators and
disregards the hidden motion. As a result, Einstein quanta of structured light are subluminal and massive.
In the semiclassical case, one should use the model of the centroid \emph{with a nonzero
kinematic (Lorentz-invariant) mass}. The analysis fulfilled unambiguously shows
that the semiclassical description is self-consistent as well as the quantum-mechanical one. For Laguerre-Gauss light beams, the
applicability of the model of massive centroid is a nontrivial consequence of
independence of centroid parameters from the longitudinal coordinate.
Properties of the Laguerre-Gauss electron are very similar.
We predict the new effect of a quantization of the velocity and mass of the
structured photon and electron. This effect is observable for the photon.
We have considered the twisted and untwisted semiclassical photon and electron (modeled by the centroids)
in the accelerated and rotating noninertial frame and have determined their \emph{inertial} masses.
Amazingly, the kinematic and inertial masses of these particles coincide. The
orbital magnetic moment of the Laguerre-Gauss electron does not depend on the
radial quantum number.

A deep similarity between fundamental properties of the
structured photon and electron illustrates the validity of statement that
the results for the photon can be well applied to all paraxial beams.

\smallbreak

\begin{acknowledgments}
This work was supported by the Belarusian Republican Foundation
for Fundamental Research
(Grant No. $\Phi$18D-002), by the National Natural Science
Foundation of China (Grants No. 11575254 and No. 11805242), and
by the National Key Research and Development Program of China
(No. 2016YFE0130800).
A. J. S. also acknowledges hospitality and support by the
Institute of Modern
Physics of the Chinese Academy of Sciences. The authors are
grateful to
I. P. Ivanov and O. V. Teryaev for helpful exchanges.
\end{acknowledgments}

\end{document}